\documentclass[twocolumn,prd,amsmath,amssymb,epsf]{revtex4-1}

\usepackage[dvipdfmx]{graphicx}
\usepackage{dcolumn}
\usepackage{bm}

\begin{document}


\title{Escape of cosmic rays from perpendicular shocks in the interstellar magnetic field}
\author{Shoma F. Kamijima}
\author{Yutaka Ohira}
\affiliation{Department of Earth and Planetary Science, The University of Tokyo, 7-3-1 Hongo, Bunkyo-ku, Tokyo 113-0033, Japan}


\begin{abstract}
We investigate the escape process from a perpendicular shock region of a spherical shock in the interstellar medium (ISM). 
The diffusive shock acceleration in the perpendicular shock of supernova remnants (SNRs) has been expected to accelerate cosmic rays (CRs) to the PeV scale without an upstream magnetic field amplification.  
We estimate the maximum energy of CRs limited by the escape from the perpendicular shock region.
By performing test particle simulations, we confirm the theoretical estimation, showing that the escape-limited maximum energy in the perpendicular shock is several $10~{\rm TeV}$ for the typical type Ia SNRs. 
Therefore, in order for SNRs in the ISM to accelerate CRs to the PeV scale, an upstream magnetic field amplification is needed.
The characteristic energy scale of several $10~{\rm TeV}$ could be the origin of the spectral break around $10~{\rm TeV}$, which was reported by recent direct CR observations. 
In addition, we show that, in the free expansion phase, the rapid perpendicular shock acceleration works on about 20\% area of the whole shock surface, which is larger than the size of the superluminal shock region. 
We also discuss the escape of CR electrons from the perpendicular shock.
\end{abstract}

\maketitle
\section{Introduction} \label{sec:intro}
It is believed that Galactic cosmic rays (CRs) are accelerated by the diffusive shock acceleration (DSA) in supernova remnants (SNRs) \cite{bell78,cr}.
In DSA, the diffusive particles can cross the shock front many times by the diffusive motion, so that they are accelerated by numerous shock compressions.
The break around PeV in the energy spectrum of observed CRs suggests that SNRs accelerate particles up to PeV. 
HESS found the evidence of a PeVatron near the Galactic center \cite{hess}.
HAWC reported some candidates of PeVatrons in our galaxy \cite{hawc}.
Recently, the Tibet AS$\gamma$ experiment found a potential PeVatron supernova remnant in our galaxy and observed sub-PeV diffuse gamma rays from the Galactic disk \cite{tibet}. 
In addition, the LHAASO recently reported some PeVatron candidates in our galaxy \cite{lhaaso}. 
In the near future, more Galactic PeVatron candidates would be detected by experiments at the Southern hemisphere (ALPACA \cite{alpaca} and SWGO \cite{swgo}). 
However, we have not understood what type of SNRs accelerates CRs to PeV. 
It was claimed that DSA in SNRs cannot accelerate CRs to PeV in the magnetic field of the interstellar medium (ISM), which is about a few ${\rm \mu G}$ \cite{cesarsky81}. 
The magnetic field must be amplified to about $100 {\rm \mu G}$ in the shock upstream region to accelerate CRs to PeV.  
Although many mechanisms for the magnetic field amplification have been proposed and investigated by several simulations \cite{lucek00,bell04,crsi}, we still do not know whether or not the sufficient magnetic field amplification in the upstream region can be realized. 
A pulsar wind nebula inside the SNR could reaccelerate CRs to the PeV scale without the upstream magnetic field amplification \cite{ohira18}.

On the other hand, it was proposed that CRs can be rapidly accelerated to PeV by DSA in the perpendicular shock of SNRs without the magnetic field amplification \cite{cesarsky81, jokipii87}. 
The rapid acceleration at perpendicular shocks was confirmed by numerical simulations \cite{rapidperp}.
Afterward, it was claimed that the momentum spectrum of accelerated particles in the perpendicular shocks becomes steeper than the canonical spectrum, $dN/dp \propto p^{-2}$, when a weak magnetic turbulence is assumed both in the upstream and downstream regions to achieve the rapid acceleration \cite{takamoto15}.
Since the downstream strong turbulence is expected by observations and simulations \cite{bamp,ohira17}, 
we recently considered weak and strong magnetic turbulences in the upstream and downstream regions, respectively, realizing the rapid acceleration and the canonical spectrum simultaneously \cite{kamijima20}. 
However, it is not understood how large area in the whole SNR surface is covered by the rapid perpendicular shock acceleration region.
The angular resolution of current gamma-ray observations is too low to identify where CR protons are accelerated in SNRs, in particular for gamma rays above 100 TeV.

\begin{figure}
\centering	
\includegraphics[scale=0.41]{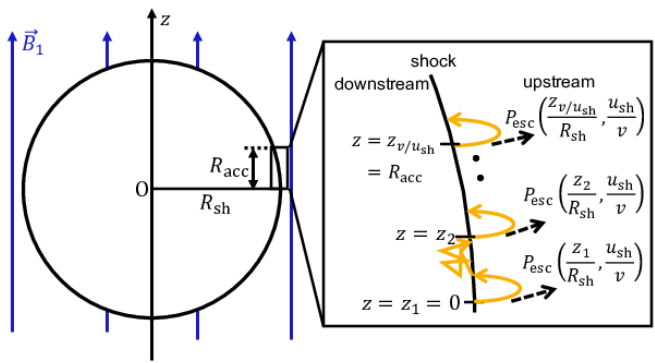}
\caption{
Schematic picture of a spherical shock wave in the ISM magnetic field and the enlarged picture of the acceleration region.
The black circle and the blue arrows are the spherical shock and the uniform magnetic field, respectively.
Yellow arrows show the schematic picture of the particle trajectory: the gyration in the upstream region and the Bohm diffusion in the downstream region.
The black dashed arrows represent the escape to the upstream region along the magnetic field line, where the escape probability depends on the distance from the equatorial plane and the shock velocity. 
\label{fig:shock}}
\end{figure}

Previous studies assumed that the maximum energy is limited by a finite age of SNRs \cite{jokipii87}. 
Whereas, it is pointed out that the maximum energy of CRs can be limited by the escape from accelerators \cite{ptuskin03}.  
Actually, spectra and images of gamma rays from middle-aged SNRs suggest that CRs are escaping from the SNRs \cite{ohira11}. 
In addition, the time evolution of the escape-limited maximum energy decides the spectrum of escaping CRs \cite{ptuskin05, ohira10}. 
Therefore, investigating the escape process and the escape-limited maximum energy are very important to understand the CR spectrum in our Galaxy. 
To investigate the maximum energy limited by the escape process, some previous studies assumed the diffusion approximation or isotropic scattering instead of solving the gyro motion in the magnetic field. 
However, the gyro motion in the upstream region has to be directly solved to reproduce the rapid acceleration in perpendicular shocks \cite{kamijima20}.
Other previous studies overcame this problem by exactly solving the gyration, but the plane shock approximation and the escape boundary condition were imposed \cite{ellison96}.
Owing to these approximation and boundary condition, the previous studies cannot correctly investigate the escape process from the perpendicular shock region on a spherical shock surface.
In this work, for a spherical non-relativistic shock wave in the ISM magnetic field (see Fig.~\ref{fig:shock}), 
we first investigate the escape process from perpendicular shocks and the time evolution of the escape-limited maximum energy.

In this work, we consider only type Ia SNRs in the ISM because most type Ia supernovae explode in the ISM, but other types of supernovae explode in the circumstellar medium (CSM). 
Although core collapse SNRs propagate to the ISM in the later phase $(t \gtrsim 10^4 ~ {\rm yr})$, where $t$ is the SNR age, the radiative cooling in the shocked ISM region becomes significant \cite{truelove99} and the velocity of the forward shock becomes too slow to accelerate CRs to the PeV scale.

In this paper, we investigate the escape process from the perpendicular shock region of the spherical shock surrounded by the ISM. 
In Sec.~\ref{sec:theory}, we show the escape process from the perpendicular shock region and derive the theoretical estimation of the size of the acceleration region and the escape-limited maximum energy.
The setup of test particle simulations and simulation results are presented in Sec.~\ref{sec:simulation}. 
Sec.~\ref{sec:discussion} is denoted to a discussion.
We summarize in Sec.~\ref{sec:summary}.


\section{Theory} \label{sec:theory}

\subsection{Parameters of type Ia supernova remnants} \label{subsec:parameter}

In this work, we use the analytical formula as the time evolution of the shock velocity of type Ia SNRs \cite{mckee95}.
The shock velocity of SNRs, $u_{\rm sh}$, is almost constant in the free expansion phase ($t \le t_{\rm ST}$) and proportional to $t^{-3/5}$ in the adiabatic expansion phase ($t \ge t_{\rm ST}$), where $t_{\rm ST} \sim 200~{\rm yr}$ is the transition time from the free expansion phase to the adiabatic expansion phase \cite{st, mckee95}.
Here, we assume the explosion energy of supernovae, $E_{\rm SN}=10^{51}~{\rm erg}$, the ejecta mass, $M_{\rm ej}=1\ M_{\odot}$,  and the mass density of the ISM, $\rho =1.67 \times 10^{-24}~ {\rm g \ cm}^{-3}$. 
The time evolution of the shock radius, $R_{\rm sh}$, is proportional to $t$ in the free expansion phase and $t^{2/5}$ in the adiabatic expansion phase. 

In our theoretical study, the upstream magnetic field, $B_1$, consists of only the uniform magnetic field component. The upstream magnetic field strength is set to be $B_1=3~{\rm \mu G}$. 
On the other hand, the downstream magnetic field, $B_2$, is highly turbulent.
Hereafter, we use subscripts 1 and 2 as the upstream and downstream values, respectively.
In this work, the magnetic field strength in the downstream region is assumed to be amplified from the shock compressed value. 
We assumed that a fraction of the energy flux of the upstream kinetic energy is converted to the downstream magnetic energy flux.
Then, the downstream magnetic field strength is 
\begin{eqnarray}
B_2 = \sqrt {4 \pi r \varepsilon_B \rho} u_{\rm sh}~~, 
\label{eq:Bdown}
\end{eqnarray}
where $\varepsilon_B$ is the conversion fraction, $r=u_1/u_2$ is the shock compression ratio, $u_1$ and $u_2$ are the upstream and downstream plasma flow velocities in the shock rest frame, and $u_{\rm sh}=u_1$ is the shock velocity in the ISM rest frame. 
We adopt $\varepsilon_B=10^{-2}$ as the fiducial value in this work.

Our magnetic field model is consistent with observations of synchrotron radiation from SNRs. 
Synchrotron radiation from the shock upstream region has never been identified in any SNRs. 
In our model, high-energy particles can propagate only their gyroradius to the shock upstream region, which cannot be resolved by any current instruments \cite{bamba03}. 
On the other hand, synchrotron radiation from shock downstream regions of SNRs has been observed. 
Those results strongly suggest that the downstream magnetic field is strongly amplified \cite{berezhko03}, which is consistent with our model.
As for the downstream magnetic field amplification, MHD and hybrid simulations show that the downstream magnetic field is strongly amplified by the downstream turbulence that is generated by an interaction between density fluctuations and a shock front \cite{bamp}.
There must be density fluctuations in the interstellar medium. In addition, there are some instabilities that can amplify the density fluctuations, the Drury instability \cite{drury86}, the instability induced by the ionization in the upstream region of leaking hydrogen atoms from the downstream region \cite{ohira13,ohira14}, and the non-linear effect of the Bell instability \cite{bell04}.
The spatial scale of the magnetic field amplification is comparable to that of the density fluctuation \cite{fraschetti13}.
In this work, we assume the sharp boundary between $B_1$ and $B_2$.
As long as the thickness of the boundary between $B_1$ and $B_2$ is much smaller than the gyroradius of high-energy particles, the sharp boundary approximation is valid.

\subsection{Acceleration time}
We theoretically estimate the escape-limited maximum energy in the perpendicular shock. 
It is estimated by equating the acceleration time of DSA, $t_{\rm acc}$, to the escape time, $t_{\rm esc}$. 
The acceleration time of DSA, $t_{\rm acc} = p \Delta t/\Delta p$, is given by the mean momentum gain per one cycle, $\Delta p/p=(4/3) (u_1 - u_2)/v$ \cite{bell78}, and the mean one cycle time, $\Delta t$ \cite{drury83}, where $v$ is the particle velocity. 
The mean one cycle time, $\Delta t$, is represented by the sum of the upstream residence time, $\Delta t_1$, and the downstream residence time, $\Delta t_2$. 
To realize the rapid acceleration and the canonical spectrum of $dN/dp\propto p^{-2}$ simultaneously, we consider weak and strong magnetic field fluctuations in the upstream and downstream regions, respectively \cite{kamijima20}. 
In that case, the upstream residence time, $\Delta t_1$, is a half of the gyro period, $\pi \Omega_{\rm g,1}^{-1}$, and the downstream residence time, $\Delta t_2$, is given by $4 \kappa_2/(u_2 v)$, where $\Omega_{\rm g,1}$ is the upstream gyro angular frequency, and $\kappa_2 = \Omega_{\rm g,2}^{-1} v^2/3$ is the downstream diffusion coefficient in the Bohm limit. 
For $\varepsilon_B=10^{-2}$, the upstream residence time, $\Delta t_1$, is longer than the downstream one, $\Delta t_2$, for $t \lesssim 2t_{\rm ST}$. 
The acceleration time, $t_{\rm acc}$, is given by \cite{kamijima20}
\begin{eqnarray}
t_{\rm acc} &=& \frac{3 r}{4 (r -1)} \left( \frac{u_{\rm sh}}{v} \right)^{-2} \left( \frac{B_2}{B_1} \right)^{-1} \nonumber \\
&&\times \left\{ \pi \left( \frac{u_{\rm sh}}{v} \right) \left( \frac{B_2}{B_1} \right) + \frac{4 r}{3} \right\} \Omega_{\rm g,1}^{-1}~~.
\label{eq:tacc}
\end{eqnarray}
The above equation was confirmed by numerical simulations of DSA at the perpendicular shock  \cite{kamijima20}. 
The ratio of two residence times, $\Delta t_1/\Delta t_2$, becomes larger when the shock velocity is fast or the downstream magnetic field strength is strong.

\subsection{Escape time and diffusion coefficient}
Next, we estimate the escape time scale from the perpendicular shock region where CRs are rapidly accelerated. 
For simplicity, we consider a spherical shock in a uniform magnetic field (see Fig.~\ref{fig:shock}). 
The $z$ axis is parallel to the uniform magnetic field. 
The origin of the coordinate is set to be the center of the spherical shock.

In our model, although particles are not scattered by the magnetic field fluctuations in the upstream region, particles are scattered in the downstream region.
Therefore, when we consider the particle motion in a timescale longer than the one cycle time, $\Delta t_1 + \Delta t_2$, the particle motion along the upstream uniform magnetic field can be regarded as diffusion.
In this work, we can assume that the shock velocity and the shock radius of SNRs are constant because the escape time is much smaller than the dynamical timescale of SNRs.
Since the diffusion coefficient linearly increases with the particle energy (see Eq.~(\ref{eq:kzz})) and the particle energy linearly increases with time, the diffusion coefficient of accelerating particles linearly increases with time.
The spacial distribution along the $z$ direction of accelerating particles, $g(z,t)$, is given by the one-dimensional diffusion equation in the case where the diffusion coefficient is proportional to time.
Then, the solution is given by
\begin{equation}
g(z,t) = \frac{N}{\sqrt{2 \pi \kappa_{zz} t }} \exp{\left[ - \frac{z^2}{2 \kappa_{zz} t} \right]}
\label{eq:sol_diffeq}
\end{equation}
where $N$ and $\kappa_{zz}$ is the number of particles and the diffusion coefficient along the upstream magnetic field line, respectively.
From the relation between the diffusion lengthscale and time, $z^2 \sim 2\kappa_{zz} t$, the escape time, $t_{\rm esc}$, can be evaluated by using the size of the acceleration region, $R_{\rm acc}$, as follows: 
\begin{equation}
t_{\rm esc} = \frac{R_{\rm acc}^2}{2 \kappa_{zz}}~~.
\label{eq:tesc1}
\end{equation}

Next, we estimate the diffusion coefficient, $\kappa_{zz}$. 
The scattering time becomes the one cycle time, $\Delta t = \Delta t_1 + \Delta t_2$. 
The mean square of the displacement along the upstream magnetic field line for $\Delta t$ is given by $(\Delta z)^2 = (\Delta z_1)^2 + (\Delta z_2)^2$, where $\Delta z_1$ and $\Delta z_2$ are displacements for $\Delta t_1$ and $\Delta t_2$, respectively.
Since particles freely move along the magnetic field line in the upstream region, the displacement in the upstream region is estimated by 
\begin{eqnarray}
\Delta z_1 = \sqrt{\langle v_z ^2\rangle} \Delta t_1 = \frac{\pi r_{\rm g,1}}{2}~~,
\label{eq:mfp1}
\end{eqnarray}
where $\sqrt{\langle v_z ^2\rangle} = v/2$ is used. The root mean square of the $z$ component of the velocity is calculated by using the upstream momentum distribution. The distribution is proportional to the particle flux that crosses the shock front from the downstream region to the upstream region.
Since the particle motion is diffusion in the downstream region, the displacement in the downstream region is given by 
\begin{eqnarray}
\Delta z_2 = \sqrt{2 \kappa_2 \Delta t_2} = \frac{ \sqrt{8 r} }{3} \left( \frac{u_{\rm sh}}{v} \right)^{-1/2} \left( \frac{B_2}{B_1} \right)^{-1}  r_{\rm g,1}~.
\label{eq:mfp2}
\end{eqnarray}
For $\varepsilon_B=10^{-2}$, the upstream displacement, $\Delta z_1$, is larger than the  downstream one, $\Delta z_2$, for $t \lesssim 20t_{\rm ST}$. 
From Eqs.~(\ref{eq:mfp1}) and (\ref{eq:mfp2}), the diffusion coefficient along the $z$ direction becomes 
\begin{eqnarray}
\kappa_{zz} = \frac{(\Delta z)^2}{2 \Delta t} 
&=& \left( \frac{B_2}{B_1} \right)^{-1} \left\{ \frac{\pi^2}{8} \left( \frac{u_{\rm sh}}{v} \right)  \left( \frac{B_2}{B_1} \right)^{2} + \frac{4 r}{9}  \right\} \nonumber \\
&&   \times  \left\{ \pi \left( \frac{u_{\rm sh}}{v} \right) \left( \frac{B_2}{B_1} \right) + \frac{4 r}{3} \right\}^{-1} r_{\rm g,1} v~.
\label{eq:kzz}
\end{eqnarray}
As already mentioned, the energy of accelerating particles linearly increases with time from the injection time, $t_{\rm inj}$, that is, $r_{\rm g,1} \propto t-t_{\rm inj}$, but the shock velocity of the SNR can be regarded as a constant 
because of $t-t_{\rm inj} \ll t_{\rm inj}$, where $t_{\rm inj}$ can be interpreted as
the dynamical timescale of SNRs. Hence, the diffusion coefficient of accelerating particles linearly increases with time from the injection time.

\subsection{Size of the acceleration region}
Next, we consider the size of the perpendicular shock acceleration region. 
The shock is classified by the angle, $\theta_{\rm Bn}$, between the magnetic field and the shock normal direction into two types: superluminal shocks and subluminal shocks \cite{hoffmann50}. 
The angle, $\theta_{\rm Bn}$, on the spherical shock depends only on $|z|/R_{\rm sh}$ (see Fig.~\ref{fig:shock}). 
Particles in the superluminal shock cannot escape to the far upstream region.
On the other hand, in the subluminal shock, particles with a velocity nearly parallel to the magnetic field line can escape to the far upstream region. 

Let $P_{\rm esc}$ be the escape probability that particles escape to the upstream region during the one cycle of DSA, which depends on $|z|/R_{\rm sh}$ and $u_{\rm sh}/v$. 
Then, the probability that upstream particles can return to the downstream region during the one cycle of DSA is $1 - P_{\rm esc}$. 
To accelerate particles (to make the particle energy twice) by DSA, particles have to experience $v/u_{\rm sh}$ times shock crossings. 
The size of the acceleration region is obtained by the following condition, 
\begin{eqnarray}
P_{\rm acc} = \prod_{i = 1}^{v/u_{\rm sh}} \left[  1 - P_{\rm esc} \left( \frac{|z_i|}{R_{\rm sh}}. \frac{u_{\rm sh}}{v} \right) \right] \approx 0~~,
\label{eq:pacc_exact}
\end{eqnarray}
where $P_{\rm acc}$ is the acceleration probability that particles do not escape to the upstream region during $v/u_{\rm sh}$ cycles. 
$z_i$ is the $z$ component of the particle position when the particle experiences the $i$-th shock crossing, and $v, R_{\rm sh}$, and $u_{\rm sh}$ are assumed to be constant. 
By solving Eq. (\ref{eq:pacc_exact}) for $|z_{v/u_{\rm sh}}|$, in principle, we can estimate the size of the acceleration region, $R_{\rm acc}$, but it need a complicated calculation.

By assuming that the particles are $z=R_{\rm acc}$ during the acceleration by DSA, 
we approximate the condition, Eq.~(\ref{eq:pacc_exact}), as follows:
\begin{equation}
P_{\rm acc} \approx \left[  1 - P_{\rm esc} \left( \frac{R_{\rm acc}}{R_{\rm sh}}, \frac{u_{\rm sh}}{v} \right) \right]^{v/u_{\rm sh}} \approx 0 ~~.
\label{eq:pacc_approx}
\end{equation}
Since the escape probability, $P_{\rm esc}$, is expected to be much smaller than unity, the above condition can be reduced to a very simple form, 
\begin{equation}
P_{\rm esc} \left( \frac{R_{\rm acc}}{R_{\rm sh}}, \frac{u_{\rm sh}}{v} \right) = \frac{u_{\rm sh}}{v}.
\label{eq:pacc_approx2}
\end{equation}

\begin{figure}
\centering
\includegraphics[scale=0.38]{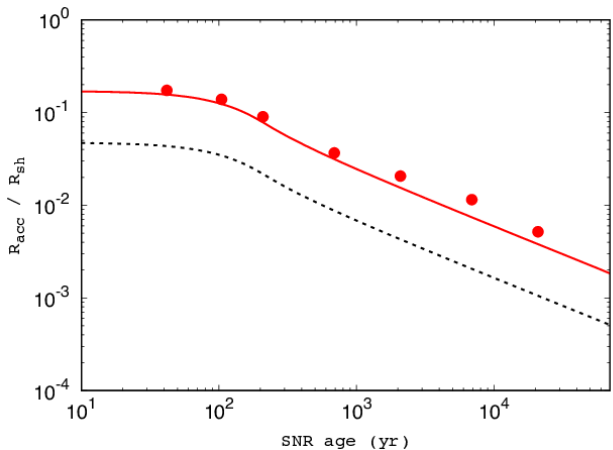}
\caption{
Size of the perpendicular shock acceleration region, $R_{\rm acc}$. 
The red solid line, red circles, and the black dotted line are the theoretical estimation (Eq.~(\ref{eq:racc_approx}), the simulation results for the uniform ISM magnetic field and the size of the superluminal shock region, respectively.
\label{fig:racc}} 
\end{figure}

Next, we numerically calculate the escape probability, $P_{\rm esc} ( |z|/R_{\rm sh}, u_{\rm sh}/v)$.
We prepare a plane shock in the uniform magnetic field and relativistic particles with $v=c$. 
The relation between $\theta_{B \rm n}$ and $|z|/R_{\rm sh}$ in the spherical shock is $\cos{\theta_{B \rm n}} = |z|/R_{\rm sh}$.
The shock velocity at $t_{\rm inj}$ is given by \cite{mckee95}.
$N_{\rm inj}=5.2 \times 10^5$ particles are injected on the shock surface. 
The momentum distribution of the injected particles is proportional to the particle flux that crosses the shock surface from the downstream region to the upstream region. 
By solving the particle trajectory for one gyro period and counting the number of particles that do not enter the downstream region, $N_{\rm esc}$, the escape probability, $P_{\rm esc}$, is calculated by $ N_{\rm esc}/N_{\rm inj}$. 
\begin{figure}
\centering
\includegraphics[scale=0.35]{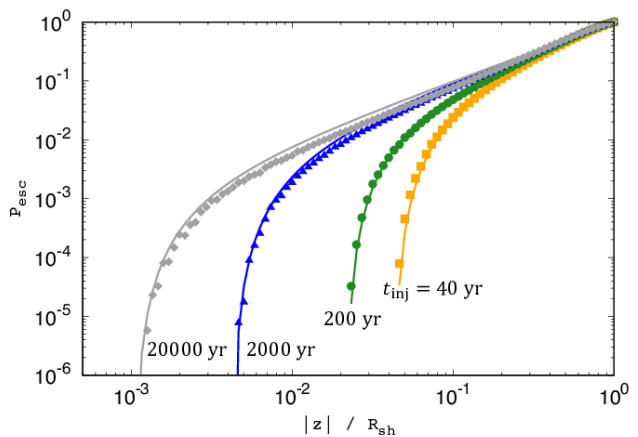}
\caption{
Escape probability that particles escape to the upstream region during the one cycle of DSA, $P_{\rm esc}$, as a function of $|z|/R_{\rm sh}$.
The orange squares, green circles, blue triangles, and grey diamonds are numerical results for $t_{\rm inj} = 40, 200, 2000$, and $20000~{\rm yr}$, respectively.
The solid lines show approximate formulae for each $t_{\rm inj}$.
\label{fig:pesc_z} }
\end{figure}
Fig.~\ref{fig:pesc_z} shows numerical results and approximate formulae (Eq.~(\ref{eq:fit_pesc})) for the escape probability, $P_{\rm esc}$, at $|z|/R_{\rm sh}$. 
The orange squares, green circles, blue triangles, and grey diamonds are numerical results for $t_{\rm inj} = 40, 200, 2000$, and $20000~{\rm yr}$, respectively. 
The solid lines show approximate formulae for each $t_{\rm inj}$.
The escape probability is zero in the superluminal shock region $(|z|/R_{\rm sh} \lesssim u_{\rm sh}/v )$. 
Furthermore, the escape probability approaches unity toward $|z|/R_{\rm sh}=1$ (the parallel shock region). 
From these asymptotic behaviors, we found the following approximate formula of $P_{\rm esc}$,
\begin{eqnarray}
P_{\rm esc} \left( \frac{|z|}{R_{\rm sh}}, \frac{u_{\rm sh}}{v} \right) = \frac{|z|}{R_{\rm sh}} \left( 1 -  \frac{u_{\rm sh} R_{\rm sh}}{v |z|} \right)^\alpha~~,
\label{eq:fit_pesc}
\end{eqnarray}
where the best-fit value for $\alpha$ is 2.6.
From Eqs.~(\ref{eq:pacc_approx2}) and (\ref{eq:fit_pesc}), $R_{\rm acc}$ becomes
\begin{eqnarray}
R_{\rm acc} \approx  (1 + \alpha)\frac{u_{\rm sh}}{v} R_{\rm sh} \approx 3.6\frac{u_{\rm sh}}{v} R_{\rm sh}~~,
\label{eq:racc_approx}
\end{eqnarray}
where  $u_{\rm sh} R_{\rm sh}/(v R_{\rm acc}) \ll 1$ and $\alpha = 2.6$ are assumed.
The red solid line and the black dotted line in Fig.~\ref{fig:racc} show the size of the acceleration region (Eq.~(\ref{eq:racc_approx})) and the size of the superluminal shock region, respectively. 
The size of the acceleration region is always larger than the size of the superluminal shock region  ($|z|/R_{\rm sh} \approx u_{\rm sh}/v$), and occupies $20\%$ area of the whole shock surface in the free expansion phase.

In $|z| \ll R_{\rm acc}$, the escape probability in the upstream region (Eq.~(\ref{eq:fit_pesc})) is much smaller than that in the downstream region ($\approx u_{\rm sh}/v$).
Therefore, the spectral softening of accelerated particles due to escape to the upstream region is negligible, and the whole energy spectrum becomes the canonical spectrum, $dN/dp \propto p^{-2}$.
On the other hand, the escape probability at $|z| = R_{\rm acc}$ is $u_{\rm sh}/v$ (see Eq.~(\ref{eq:pacc_approx2})).
Furthermore, the upstream escape probability increases with $|z|$.
Thus, in $|z| \ge R_{\rm acc}$, the escape probability in the upstream region is larger than that in the downstream region, so that a large number of particles escape to the upstream region. 
As a result, the energy spectrum has a cutoff feature at the escape-limited maximum energy, $E_{\rm max,esc}$.

\subsection{Maximum energy}
Equating the acceleration time to the escape time, the escape-limited maximum energy in the perpendicular shock region is calculated by 
%
%
\begin{eqnarray}
E_{\rm max,esc} &=& \sqrt{\frac{4 (r - 1) (1 + \alpha)^2}{3 r}} \left( \frac{u_{\rm sh}}{v} \right)^2  \left( \frac{B_2}{B_1} \right)  \nonumber \\ 
&&\times \left\{ \frac{\pi^2}{4} \left( \frac{u_{\rm sh}}{v} \right)  \left( \frac{B_2}{B_1} \right)^{2} + \frac{8 r}{9}  \right\}^{-1/2} Z e B_1 R_{\rm sh}~,~~~~
\label{eq:emaxesc}
\end{eqnarray}
where $Ze$ is the electric charge of the accelerated particle, and we use Eq.~(\ref{eq:racc_approx}) as the size of the acceleration region. 
$u_{\rm sh}, R_{\rm sh}, B_1$, and $B_2$ are given in Sec.~\ref{subsec:parameter}.
$R_{\rm acc}$ is obtained from the results in Fig.~\ref{fig:racc}. 
We can evaluate the escape-limited maximum energy, $E_{\rm max,esc}$, by using these quantities and Eq.~(\ref{eq:emaxesc}).
The condition, $t_{\rm acc}=t_{\rm esc}$, is reduced to $p/\Delta p = R_{\rm acc}^2/(\Delta z)^2$ because the one cycle time, $\Delta t$, cancels out from the condition. 
Thus, the behavior of $E_{\rm max,esc}$ is determined by whether the displacement along the upstream magnetic field, $\Delta z$, is given by the upstream displacement, $\Delta z_1$, or the downstream displacement, $\Delta z_2$.
For $t \lesssim 20t_{\rm ST}$, $\Delta z \approx \Delta z_1$, so that the escape-limited maximum energy is calculated by 
\begin{eqnarray}
	E_{\rm max,esc} &\approx& \sqrt{\frac{16 (r - 1) (1 + \alpha)^2}{3 \pi^2 r}}  \left( \frac{u_{\rm sh}}{v} \right)^{3/2} Ze B_1 R_{\rm sh}~~~~\\
	&\propto& 
	\left\{
	\begin{array}{ll}
		t &(t \le t_{\rm ST}) \\
	 	t^{-1/2} &(t_{\rm ST} \le t \le 20t_{\rm ST}) \nonumber
	\end{array}~~.
	\right.
\end{eqnarray}
For $t \gtrsim 20t_{\rm ST}$, $\Delta z \approx \Delta z_2$, so that $E_{\rm max,esc}$ is 
\begin{eqnarray}
E_{\rm max,esc} &\approx& \sqrt{\frac{3(r - 1)(1 + \alpha)^2}{2 r^2}} \left( \frac{u_{\rm sh}}{v} \right)^2 \left( \frac{B_2}{B_1} \right) Ze B_1 R_{\rm sh} ~~~~~\\ 
&\propto&  t^{-7/5} ~~(t \ge 20t_{\rm ST})~~. \nonumber
\end{eqnarray}
The escape-limited maximum energy for protons becomes maximum at $t \approx t_{\rm ST}$, 
\begin{eqnarray}
&E_{\rm max,esc}&(t=t_{\rm ST}) = 6.1 \times 10^{13}~{\rm eV} \left( \frac{B_1}{3{\rm \mu G}} \right) \left( \frac{E_{\rm SN}}{10^{51}~{\rm erg}} \right)^{3/4}  \nonumber \\ 
&&\times \left( \frac{M_{\rm ej}}{M_\odot} \right)^{-5/12} \left( \frac{\rho}{1.67\times10^{-24}{\rm g~cm^{-3}}} \right)^{-1/3},~~~~
\end{eqnarray}
where $r=4$, $\alpha=2.6$, $v=c$, and fiducial parameters for type Ia SNRs (see Sec.~\ref{subsec:parameter}) are used.

Equating the acceleration time and the SNR age, $t_{\rm age}$, the age-limited maximum energy in the perpendicular shock is given by
\begin{eqnarray}
E_{\rm max,age} &=& \frac{4(r - 1)}{3 r} \left( \frac{u_{\rm sh}}{v} \right)^{2}  \left( \frac{B_2}{B_1} \right)  \nonumber \\
&& \times \left\{ \pi \left( \frac{u_{\rm sh}}{v} \right) \left( \frac{B_2}{B_1} \right) + \frac{4 r}{3} \right\}^{-1} Z e B_1 v t_{\rm age}~.~~~~~
\label{eq:emaxage}
\end{eqnarray}

\begin{figure}
\centering
\includegraphics[scale=0.38]{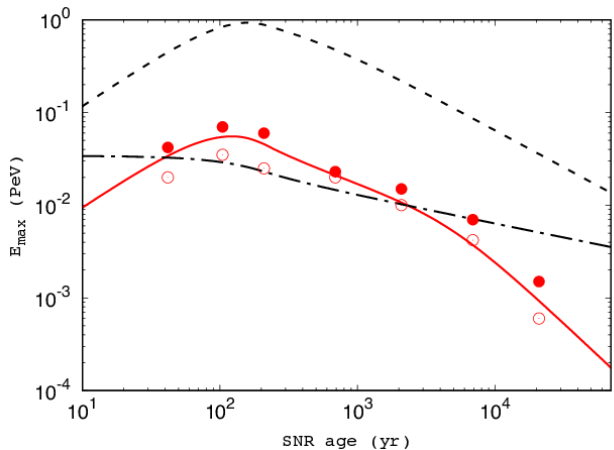}
\caption{Evolution of the maximum energy of particles accelerated in the perpendicular shock of type Ia SNRs.  The black dashed line, red solid line, and black dot-dashed line are the age-limited maximum energy (Eq.~(\ref{eq:emaxage})), the escape-limited maximum energy (Eq.~(\ref{eq:emaxesc})), and the cooling-limited maximum energy of electrons (Eq.~(\ref{eq:emaxcool})). 
The red filled and red open circles are simulation results for the uniform and turbulent magnetic field with the injection scale of $100~{\rm pc}$, respectively. 
\label{fig:emax}} 
\end{figure}
Hereafter, $v=c$, $r=4$ and $Z = 1$ are assumed.
Fig.~\ref{fig:emax} shows the time evolution of the maximum energy.
The black dashed line and red solid line are the age-limited maximum energy (Eq.~(\ref{eq:emaxage})) and the escape-limited maximum energy (Eq.~(\ref{eq:emaxesc})), respectively. 
As one can see in Fig.~\ref{fig:emax}, the escape-limited maximum energy is always at least ten times smaller than the age-limited maximum energy. Therefore, the maximum energy in the perpendicular shock region is always limited by the escape process, even for the free expansion phase. 
In this theoretical study, CRs can be accelerated to 30 TeV in the perpendicular shock region of the typical type Ia SNR in the ISM.

\section{Simulation} \label{sec:simulation}
\subsection{Simulation setup}
To confirm the above theoretical estimate, we perform test particle simulations of CR acceleration and escape processes in the perpendicular shock. 
We consider a non-relativistic spherical shock wave in the ISM magnetic field (see Fig.~\ref{fig:shock}), which corresponds to typical type Ia SNRs.
All parameters are given in Sec.~\ref{subsec:parameter}.
For the velocity profile inside the spherical shock, we use an approximate solution of $u_2(R) = (3 u_{\rm sh} /4) (R/R_{\rm sh})$, where $R$ is the radial distance from the explosion center \cite{ohira18}.

Different methods are used to solve the particle motion in the upstream and downstream regions.
We solve random work by the Monte-Carlo method in the downstream region 
because we assume that the downstream magnetic field is highly turbulent, and particles are isotropically scattered in the local fluid rest frame.
The diffusion is set to be the Bohm limit in the downstream magnetic field. 
On the other hand, in the upstream region, we directly solve the particle orbit in the upstream magnetic field, $\vec{B}_1$.
The Bunemann-Boris method is used to solve the equation of motion \cite{birdsall91}.

In this simulation study, we consider two types of the upstream magnetic field.
One is a uniform magnetic field, $\vec{B}_1=\vec{B}_0$. 
The strength of $\vec{B}_0$ is set to be $3~{\rm \mu G}$.
The other is the uniform magnetic field with a turbulent magnetic field, $\vec{B}_1=\vec{B}_0+\delta \vec{B}$. 
The magnetic field fluctuation in the upstream region, $\delta \vec{B}$, is represented by the summation of static plane waves \cite{hussein14}.
We use the isotropic Kolmogorov spectrum as the magnetic field fluctuation. 
The injection scale of turbulence is set to be 1 pc or 100 pc. 
The amplitude of the magnetic field fluctuations is set to be $|\delta \vec{B}| = |\vec{B}_0|$.

The simulation particles are impulsively injected at $t_{\rm inj}=40, 100, 200, 660, 2000, 6600,$ and $20000~{\rm yr}$. 
For the uniform magnetic field, particles are injected at the equator ($z=0$) on the spherical shock, that is, particles are injected at the perpendicular shock. 
For the fluctuated magnetic field, particles are injected on the whole shock surface.
The initial particle momentum distribution is isotropic in the momentum space. 
The injected particle energy is $100~{\rm GeV}$ for $t_{\rm ini} = 20000~{\rm yr}$ and $1~{\rm TeV}$ for the other injection times. 
We use the particle splitting method to improve statistics of the number of high-energy particles.

\subsection{Simulation results}
The simulation results for the maximum energy are plotted in Fig. \ref{fig:emax}. 
The red filled and red open circles are simulation results for the uniform and fluctuated magnetic field with the injection scale of $100~{\rm pc}$, respectively, which are extracted from the simulation results as follows. 
We first make the momentum spectrum of escaping CRs, $p^2 dN/dp$, and estimate the peak momentum at every some time steps. 
When particles move more than ten percent of the shock radius away from the shock surface, we regard the particles as escaping particles. 
Next, we see the time evolution of the peak momentum of $p^2 dN/dp$ for escaping CRs, identifying the maximum value of the peak momentum as the escape-limited maximum energy. 
As one can see, the simulation results are in agreement with the theoretical estimation about the escape-limited maximum energy. 
Even though the upstream magnetic field is assumed to be uniform in the theoretical estimation, our simulations show that the escape-limited maximum energy in the realistic model of the ISM magnetic field does not change significantly from the theoretical estimation. 
Although the results are not shown here, even for the fluctuated magnetic field with the injection scale of $1~{\rm pc}$, the results do not change significantly from that for $100~{\rm pc}$. 
Therefore, we confirmed our theoretical study that the maximum energy of particles accelerated in the perpendicular shock in the ISM is limited by the escape process. In addition, the perpendicular shock of type Ia SNRs in the ISM cannot accelerate CRs to the PeV scale. 

Simulation results about the size of the perpendicular shock region are plotted in Fig.~\ref{fig:racc}. 
The red solid line and red points are the theoretical estimation and simulation results for the uniform magnetic field, respectively. 
For the simulation, we estimated $R_{\rm acc}$ by taking the average of the $z$ coordinate of the last interaction points between the shock front and escaping particles with $E \sim E_{\rm max,esc}$.
Simulation results are in good agreement with the theoretical estimation.
The theoretical estimation about the size of the acceleration region is always slightly smaller than the simulation results $(\Delta R_{\rm acc}/R_{\rm sh} \sim 0.25)$.
This is probably due to replacing Eq.~(\ref{eq:pacc_exact}) with Eq.~(\ref{eq:pacc_approx}).
Particles injected on the shock surface at $z = 0$ diffuse to $|z| > 0$. The escape probability, $P_{\rm esc}$, increases with $|z|$ because the angle between the shock normal direction and the upstream magnetic field decreases with $|z|$. Since Eq.~(\ref{eq:pacc_approx}) approximates that particles always cross the shock front at $z=R_{\rm acc}$ until the particles escape, the escape probability is slightly overestimated. Hence, our theoretical estimation about $R_{\rm acc}$ is expected to be more consistent with the simulation results if we solve Eq.~(\ref{eq:pacc_exact}). 
In the free expansion phase, although the size of the superluminal shock region ($\sim (u_{\rm sh}/v)R_{\rm sh}$) is a few $\%$ of the whole shock surface, the rapid perpendicular shock acceleration works on about $20\%$ area of the whole shock surface.
Thus, the rapid acceleration realizes mainly in the subluminal shock region. 
The perpendicular shock region can contribute the Galactic CR production during the free expansion phase. 
Since the fraction decreases with time after the free expansion phase, the contribution to the Galactic CRs from the perpendicular shock region would decrease with time.

\section{Discussion} \label{sec:discussion}
So far, we did not consider the cooling by the synchrotron radiation, which must be taken into account to decide the maximum energy of CR electrons \cite{ohira12}.
The cooling-limited maximum energy, $E_{\rm max,cool}$, is determined by equating the energy loss by the synchrotron cooling during the one cycle time, $\Delta E_{\rm syn}$, and the energy gain during the one cycle ($\Delta E_{\rm syn} = \dot{E}_{\rm syn,1} \Delta t_1 + \dot{E}_{\rm syn,2} \Delta t_2 = (4/3) (u_1 - u_2)E/c$), where $\dot{E}_{\rm syn,1(2)} = 4 e^4 E^2 B_{1(2)}^2/(9 m_{\rm e}^4 c^7)$ \cite{kamijima20}. Then, the cooling-limited maximum energy is calculated by 
\begin{eqnarray}
E_{\rm max,cool} &=& \frac{3\sqrt{3}}{2}  \frac{(m_{\rm e} c^2)^2}{e^{3/2}}  \left( \frac{u_{\rm sh}}{c} \right)^{1/2} \nonumber \\
&&\times\left\{ 3 \pi B_1 +  64\sqrt{\pi \varepsilon_B \rho c^2} \right\}^{-1/2} \nonumber \\
&\approx& 2.70 \times 10^{13}~{\rm eV} \left( \frac{u_{\rm sh}}{0.03c} \right)^{1/2} \label{eq:emaxcool}~~,
\end{eqnarray}
where $\varepsilon_B=10^{-2}$ is assumed in the last equation. 
For $\varepsilon_B=10^{-2}$, CR electrons lose their energy mainly in the downstream region. 
During $50~{\rm yr} \lesssim t \lesssim 2000~{\rm yr}$, the cooling-limited maximum energy is smaller than the escape-limited maximum energy. 
Therefore, CR electrons can escape from the perpendicular shock region for $t \lesssim 50~{\rm yr}$ and $t\gtrsim 2000~{\rm yr}$. 
If the downstream magnetic field decreases with time more rapidly than assumed in this work, CR electrons start to escape faster than $2000~{\rm yr}$ \cite{ohira12}.

The spectral break of Galactic CR protons around $10~{\rm TeV}$ was recently reported by three experiment, CREAM, NUCLEON, and DAMPE \cite{10tev}. 
It was shown that the break energy is $16^{+13}_{-8}~{\rm TeV}$ by combining these direct CR observations \cite{lipari20}. 
Furthermore, very recently, DAMPE reported that CR heliums have a spectral break at a similar energy scale \cite{dampe}.
The origin of the spectral breaks around $10~{\rm TeV}$ is still unknown. 
As shown in Fig.~\ref{fig:emax}, the perpendicular shock region of type Ia SNRs in the ISM can naturally provide several $10~{\rm TeV}$ scale. 
To understand the origin of this break, we need further studies, especially for injection into DSA at the perpendicular shock, which is beyond the scope of this work.

After particles escape to the upstream region along a magnetic field line, the escaping particles could interact with the shock surface again thanks to the pitch angle scattering. If the scattering is not so efficient, the escaping particles are trapped in the same magnetic line for a long time. 
As the SNR shock expands, the magnetic field line interacts with the quasi-parallel shock region \cite{giacalone17}. 
As a result, particles escaped from the perpendicular shock region can affect the parallel shock region.  
Ultimately, global studies in the whole SNR system would be needed to understand CR acceleration by SNRs.

Reference \cite{giacalone17} investigated particle acceleration in a spherical shock while solving the gyration.
Compare with the early study \cite{giacalone17}, there are mainly two different points.
One is about the downstream magnetic field model.
In the early study \cite{giacalone17}, the downstream magnetic field is given by the shock compression of the upstream magnetic field.
On the other hand, in this work, we assumed that the downstream magnetic field is turbulent and amplified.
The other difference is about the particle energy.
The early study \cite{giacalone17} focused on particle acceleration in the energy range from $1~{\rm keV}$ to $10~{\rm GeV}$, and showed that low-energy particles are rapidly accelerated at the perpendicular shock region and accelerated particles move towards the parallel shock region while moving along the magnetic field line.
This behavior is consistent with this work.
On the other hand, we focused on high-energy particles from {\rm TeV} to {\rm PeV} in this work, and revealed the size of the acceleration region and the time evolution of the escape-limited maximum energy.
Therefore, this work and the early study \cite{giacalone17} are complementary.

In this work, we did not take account of CR streaming instabilities. However, in a parallel shock region, it is believed that the accelerating and escaping particles amplify the upstream magnetic field fluctuations by streaming instabilities \cite{lucek00,bell04}. 
Escaping particles from the perpendicular shock region propagate along the magnetic field line, so that the streaming instabilities could amplify the upstream magnetic field fluctuations.
If the upstream magnetic field fluctuation is amplified in a perpendicular shock region, particles can be confined for a longer time. 
Although the longer escape time would make the maximum energy larger, the upstream magnetic field fluctuations make the acceleration time in the perpendicular shock long \cite{jokipii87,takamoto15}. 
It is an interesting question whether the streaming instability by the escaping particles from the perpendicular shock region makes the maximum energy larger or not.

For core-collapse supernovae, some of the progenitor mass are stripped by the stellar wind before the explosion. 
Then, they explode in the CSM. 
The magnetic field configuration in the CSM is expected to be like the Parker spiral structure. 
Then, most regions of SNR shocks in the CSM become perpendicular shocks, so that CRs could be accelerated to the PeV scale \cite{takamoto15}. 
The polarity of the parker spiral magnetic field and the particle charge determine whether accelerating particles concentrate on the polar region or the equatorial region. 
We are going to investigate CR acceleration and escape processes in the parker spiral magnetic field in the next paper.

In this work, we assumed non-thermal high-energy particles are injected at the perpendicular shock. 
There are some injection processes at the perpendicular shock.
For perpendicular shocks in a partially ionized plasma, downstream hydrogen atoms leak into the upstream region. The leaking hydrogen atoms can be injected into DSA after they are ionized \cite{ohira16}.
Even for shocks in a fully ionized plasma, a quasi-parallel shock could be locally realized in a globally perpendicular magnetic field configuration \cite{giacalone05}.  
Thermal particles can be injected at these local parallel shocks and can be accelerated at the globally perpendicular shock.
Furthermore, if there are initially high-energy particles in the upstream region, they can be injected into DSA at the perpendicular shock \cite{caprioli18}.
Current particle-in-cell and hybrid plasma kinetic simulations cannot investigate the acceleration of TeV or PeV particles in SNRs because of a short simulation time.
Therefore, it is still unknown for the injection into DSA under the situation that TeV or PeV particles are accelerating.
In addition, the maximum energy of particles accelerated at quasi-parallel shocks is still unknown because the upstream magnetic field amplification has not been fully understood.
Therefore, it is still unknown which accelerates more particles to higher energy, parallel or perpendicular shocks.
We have to simultaneously solve injection, acceleration, escape, and the magnetic field amplification in the global system of SNRs, although it is challenging.


\section{Summary} \label{sec:summary}
In this work, we investigated the escape process of CRs accelerated in a perpendicular shock of a spherical shock in the ISM magnetic field.
The perpendicular shock in the SNRs has been expected to be the PeVatron for a long time, but the escape process of accelerated particles has never been investigated. 
Furthermore, it is not understood how large area is covered in the whole SNR surface by the rapid perpendicular shock acceleration region. 
We assumed that the particle motion is the gyration in the upstream magnetic field and the Bohm diffusion in the shock downstream region. 
If the shock surface or the upstream magnetic field have a curvature, accelerated particles escape from the perpendicular shock region by the particle motion along the upstream magnetic field line. 
The particle motion along the upstream magnetic field can be interpreted as diffusion because of the scattering in the shock downstream region. 
We showed that for the free expansion phase of SNRs, the rapid perpendicular shock acceleration is realized in about $20\%$ area of the whole SNR surface, which is larger than the size of the superluminal shock region (a few $\%$ area of the whole SNR surface).
We theoretically estimated the escape-limited maximum energy in the perpendicular shock region (Eq.~(\ref{eq:emaxesc})).
By applying the theoretical estimation to the typical type Ia SNRs and performing test particle simulations, we concluded that the maximum energy is always limited by the escape process, and the escape-limited maximum energy is about several  $10~{\rm TeV}$, that is, the perpendicular shock of SNRs in the ISM cannot be PeVatron without the magnetic field amplification.
However, the perpendicular shock of SNRs in the ISM can explain the spectral break of Galactic CR protons and helium around $10~{\rm TeV}$.
Moreover, we derived when accelerated electrons start to escape from the perpendicular shock region.

\begin{acknowledgments}
We thank M. Hoshino and T. Amano for valuable comments. 
Numerical computations were carried out on Cray XC50 at Center for Computational Astrophysics, National Astronomical Observatory of Japan. 
S.K. is supported by International Graduate Program for Excellence in Earth-Space Science (IGPEES), The University of Tokyo. 
Y.O. is supported by JSPS KAKENHI Grant Number JP19H01893, and by Leading Initiative for Excellent Young Researchers, MEXT, Japan.
\end{acknowledgments}

\end{document}